\documentstyle[editedvolume,epsf]{crckapb} 

\begin{opening}
\title{Brown Dwarfs and Very Low Mass Stars:
 Towards a New Age Scale for Young Open Clusters}

\author{D. Barrado y Navascu\'es}
\institute{Max-Planck Institut f\"ur Astronomie\\
K\"onisgtuhl 17, D-69117 Heidelberg, Germany}
\author{J.R. Stauffer}
\institute{Harvard-Smithsonian Center for Astrophysics\\
           60 Garden St. Cambridge, MA02138, USA}
\author{J. Bouvier}
\institute{Observatoire de Grenoble\\
     F-38041 Grenoble, Cedex 9, France}

\end{opening}

\runningtitle{New Age Scale for Young Open Clusters}

\begin{document}

\section{Introduction}

\begin{figure}	
\vspace{-2.0cm}
\hbox{\hspace{-0.6cm}\epsfxsize=5.8cm \epsfbox{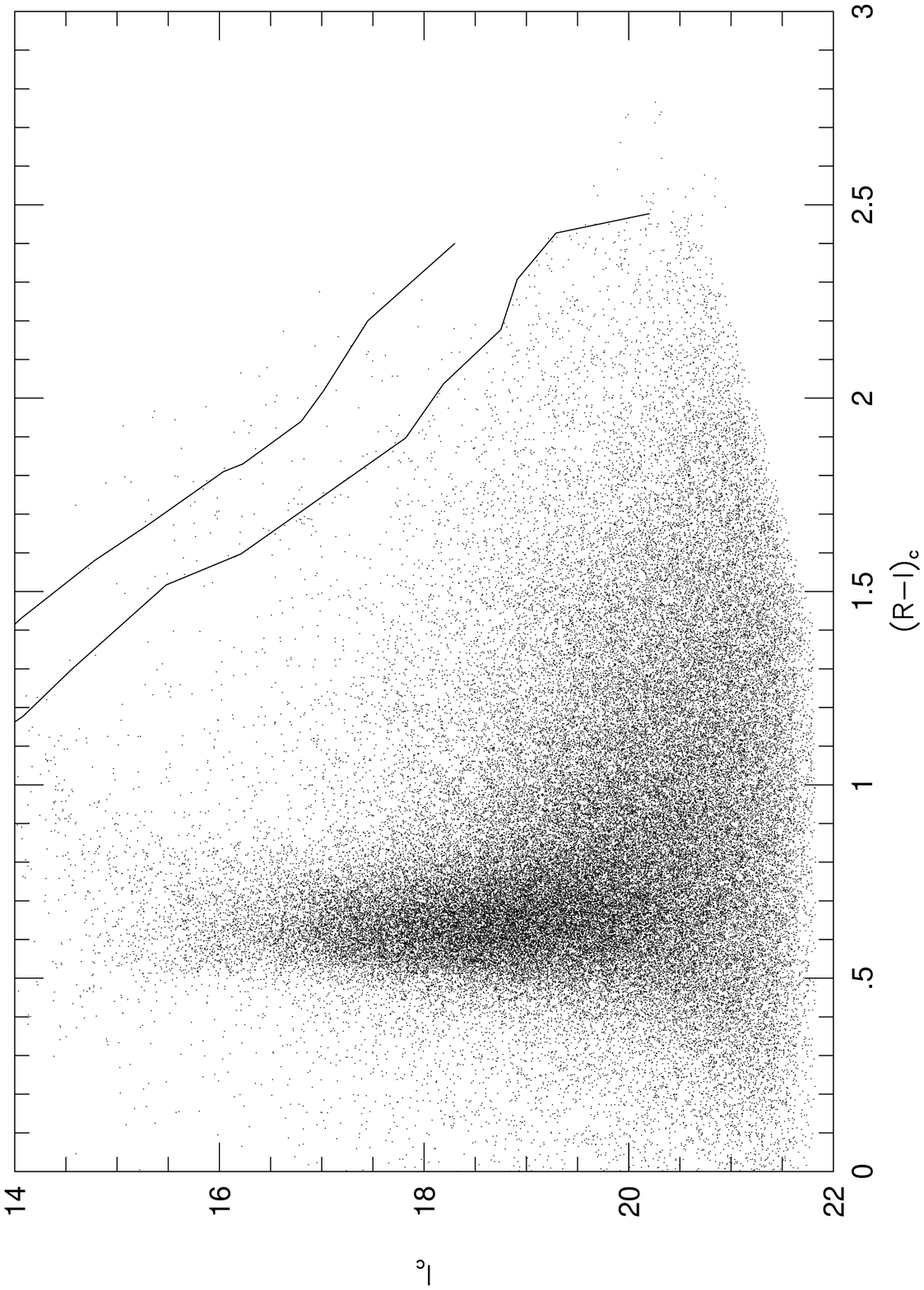}
\hspace{6.5mm}\epsfxsize=5.8cm \epsfbox{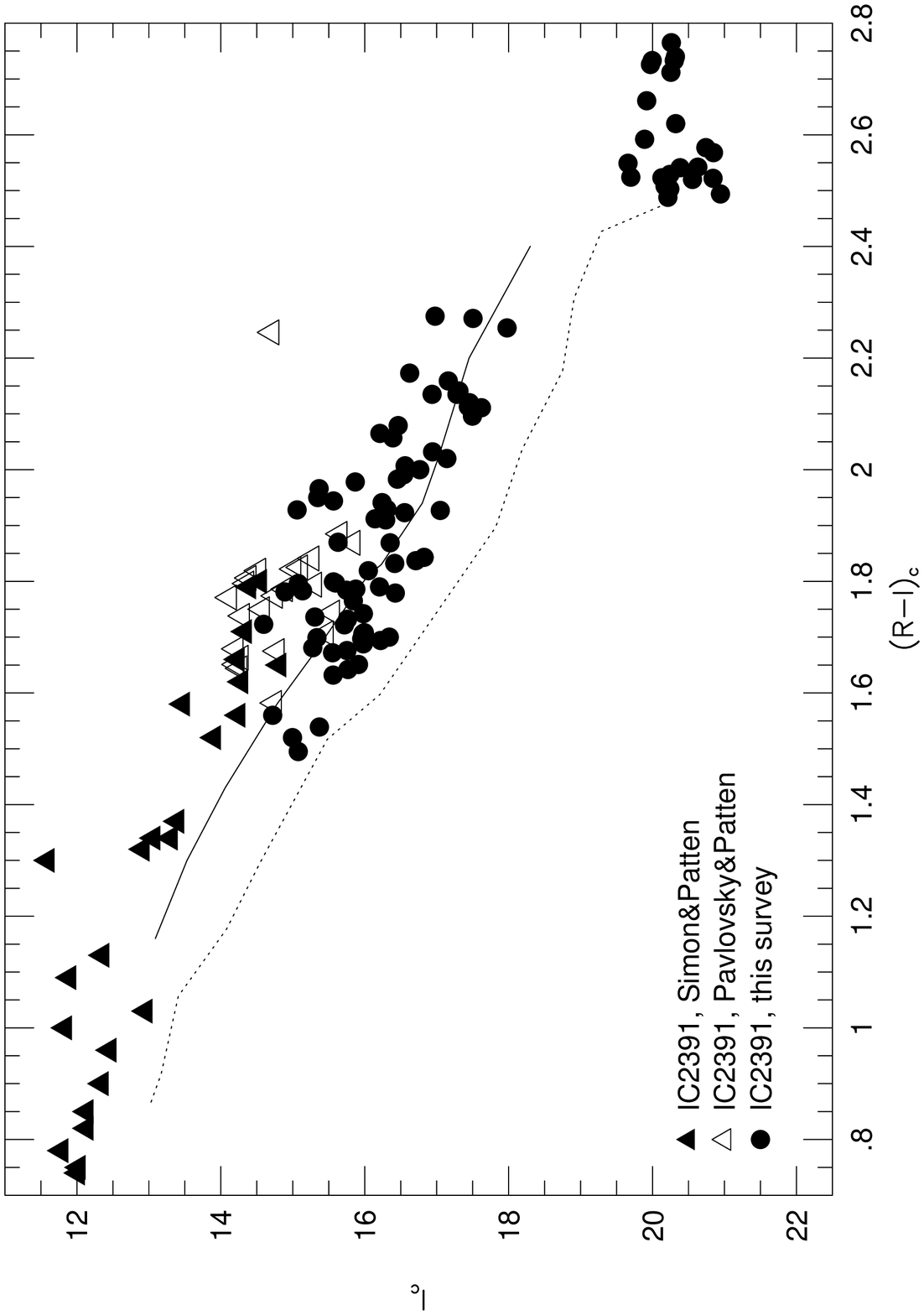}}
\vspace{-2mm}
  \caption{
{\bf a} Color-Magnitude Diagram for the area around IC~2391.  
{\bf b} Our selection of possible members of IC~2391.
}
\end{figure}

Open clusters are one of the main available laboratories in Astronomy.
Among other things, the stellar evolutionary age scale  relies primarily 
on them, so it is extremely important to estimate accurately
the ages of different associations. Until now, the dating of  clusters
was based on  either  the Upper Main Sequence (MS), using stars evolving
off it or already well beyond the turn--off point, or isochrone
fitting. These techniques
are model dependent and provide different ages for the same cluster.
Very Low Mass Stars (VLM) and
 Brown Dwarfs (BDs) offer an alternative: It is well known that the lithium
abundance is depleted with the age in stars
(see the review by Balachandran 1994). 
 This destruction depends also on mass,
so dM stars deplete their lithium even before they arrive to the MS.
However, the internal part  of VLM/BDs  is not
 hot enough to destroy  lithium at this stage.
In the lower part of the cluster MS,
 the border between members having no lithium and
those showing it in their atmosphere is
 called the lithium depletion boundary (LDB). 
As a cluster gets older, the LDB moves toward
 less massive objects.
Therefore, the detection of lithium in the brightest (e.g. bluer, hotter
and more massive than others) BDs of a cluster 
puts a very strong limit on its age (D'Antona \& Mazzitelli 1994;
Mart\'{\i}n \& Montes 1997; Ventura et al. 1998). 
In this paper, we show how this
technique works and apply it to several young open clusters.

\section{Identification of candidate members: Optical and IR surveys}

The initial step in order to identify new very low mass (VLM) stars and
BDs has been to obtain   optical photometry on the area around 
the cluster. In order to
 cover a significat fraction of a particular cluster
in a reasonable amount of time, we have used in most of the cases
detectors with several CCDs. Our targets have been:\\
i) The Pleiades.- We have used the CFHT (with the MOSAIC camera)
  and 48" telescope at MHO. We surveyed  2.5 and 1 sq. deg., respectively, 
discovering  17 and 6 BDs  candidates in each survey (Bouvier et al. 1998;
Stauffer et al. 1998a). \\
ii) $\alpha$ Per.- As a preliminary survey, we collected RI photometry
with the 48" telescope at MHO. We discovered several candidates. Some
of them were the targets of a spectroscopic campaign (next section).\\
iii) IC~2391.- The CTIO 4m telescope and the BTC camera allowed us to
survey  2 sq.  deg., detecting $\sim$40 BDs candidates (Barrado y Navascu\'es
et al. 1999a). See Fig. 1a.  \\
iv) M35.- CFHT$+$MOSAIC: This is a very well populated open  cluster.
Our study included several thousands candidates (stars) and few possible BD
(Barrado y Navascu\'es et al. 1999b).\\
v) NGC~2516.- CTIO 4m telescope  $+$ BTC. We covered 0.6  sq. deg.,
 analysis in progress.

\begin{figure}	
\vspace{-2.0cm}
\hbox{\hspace{-0.6cm}\epsfxsize=5.8cm \epsfbox{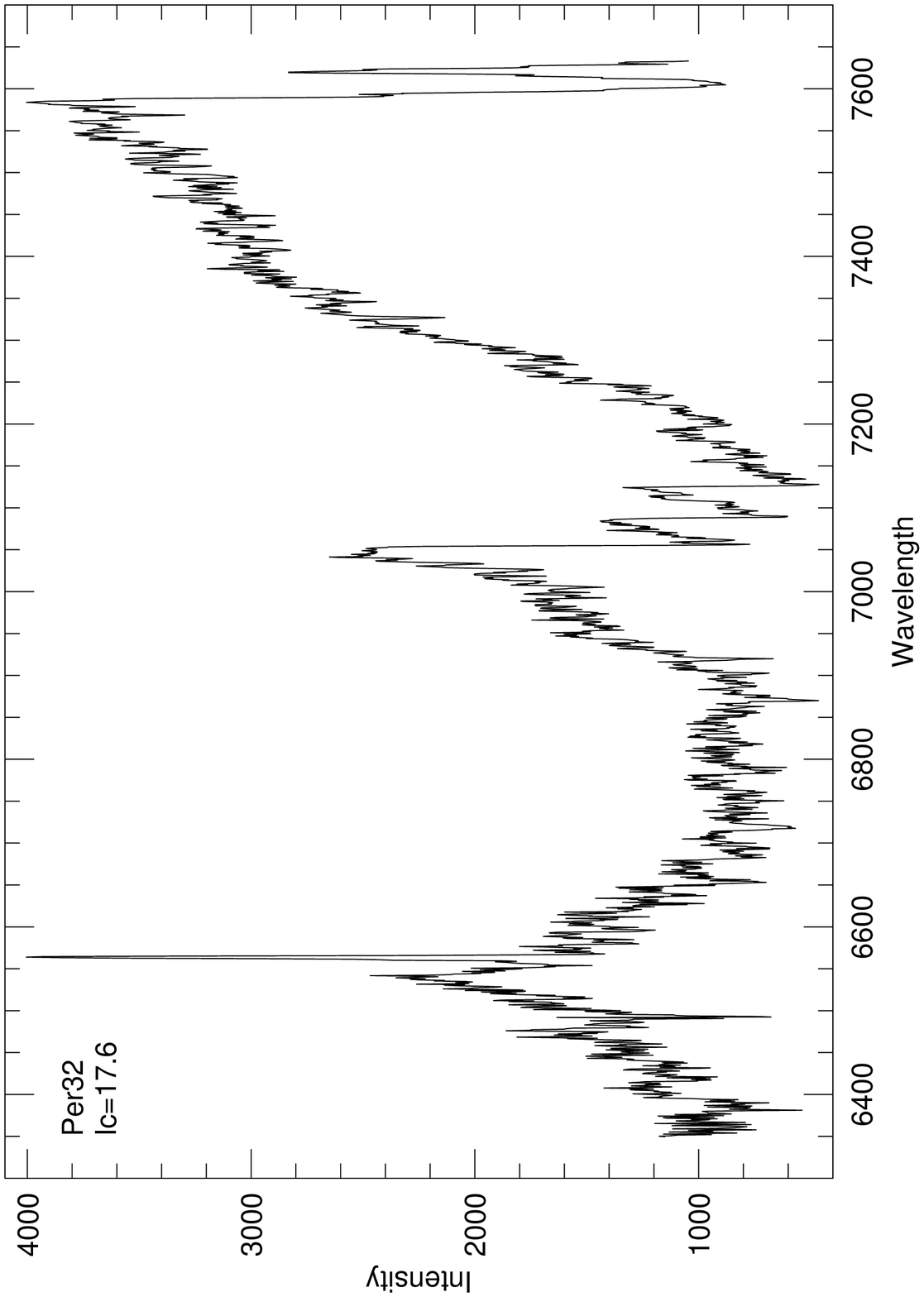}
\hspace{6.5mm}\epsfxsize=5.8cm \epsfbox{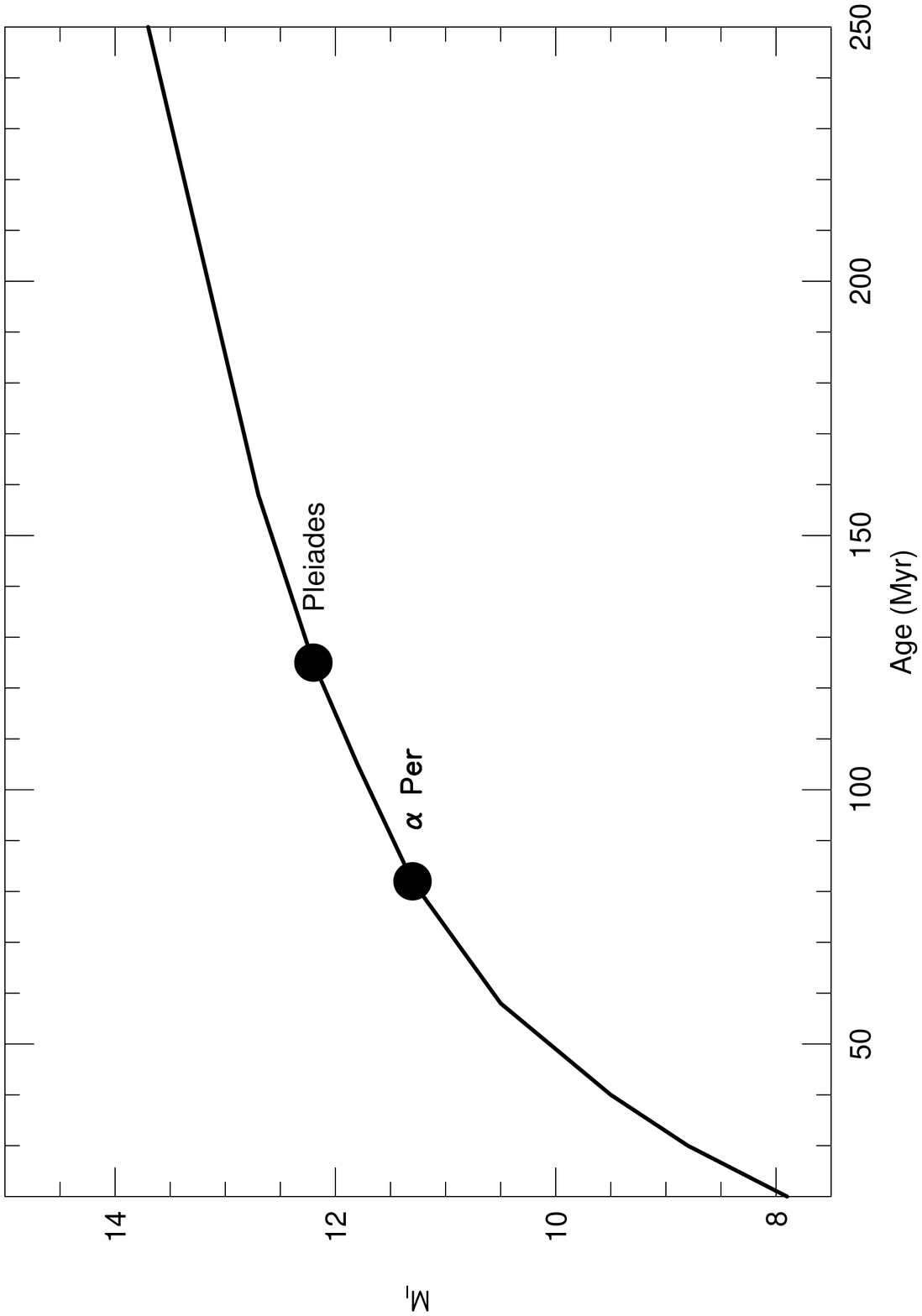}}
\vspace{-2mm}
  \caption{
{\bf a} Spectrum of a new member of $\alpha$ Per. 
{\bf b} M$_{\rm I}$ versus age. The location of the lithium
depletion boundary is shown for the Pleiades and $\alpha$ Per.
}
\end{figure}

Since most of the lists of cluster candidates contain spurious members,
it is necessary to select good candidates by obtaining near IR photometry.
We have collected this information using the MHO 48'' telescope and the
  NASA IRTF, in the case of the Pleiades. JHK photometry on the area
around IC~2391 was obtained by the   2MASS  project. All this information
allowed us to construct robust lists of possible  members. 
 Figure 1b displays a Color-Magnitude diagram of IC~2391, and a comparison 
with previous surveys. A 30  Myr isochrone (solid line) and a Zero
Age MS (dashed line) are included.

\section{Spectroscopy: membership and the new age}
 
Since these objects are too faint to have been detected with 
photographic plates in the past, most of them do not have measurements
of their proper motions. Therefore,  the final confirmation of membership 
comes from the spectroscopy. A intermediate resolution spectra 
(R=2000--5000) are good enough to obtain rough radial velocities (hence, 
establishing whether the target is a member). On the other hand, a detection
of Li{\sc I} 6708 \AA{ } reveals the BD nature of the candidate and also
confirms the membership (Rebolo 1991). 
Using the Keck II  telescope and the LRIS
spectrograph we have collected spectra of VLM and BD candidates in the 
Pleiades and $\alpha$ Per open clusters. Figure 2a shows the spectrum of
PER 32, a BD candidate discovered in the $\alpha$ Per region
 with the 48'' MHO. The intense H$\alpha$ line in emission and its radial 
velocity indicates that it is a real member of the cluster. However, we have
not detected lithium, which indicates that this is a VLM star close to the
LDB.

\begin{figure}	
\vspace{-7.5cm}
\hspace{-2cm}\epsfxsize=12.4cm \epsfbox{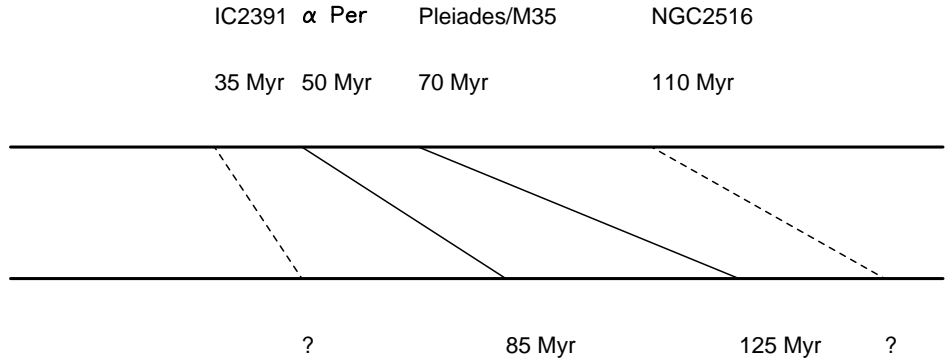}
\vspace{-2.7cm}
  \caption{Comparison between the standard age scale (top)  and the values
estimated from the lithium depletion boundary for BD (bottom).}
\end{figure}

Figure 2b illustrates how ages are estimated. Following Stauffer et al. 
(1998b), the faintest member without lithium and the brightest member
with it, bracket the magnitude of the LDB.  We used the Chabrier \& Baraffe
(1998) 
latest evolutionary model  to covert these magnitudes, different for the
 Pleiades and $\alpha$ Per,  into their corresponding ages.
 If the sampling is well done, the accuracy is better than 10\%.

Figure 3 shows the old standard ages for the 5 clusters included in our
 study, together the new values for the Pleiades and $\alpha$ Per, which
are significantly older than the values previously accepted. Our aim is to 
improve these values and to get new ones for the other clusters in the near
future, in order to verify this interesting possibility. This would have
a very important effect on several fields, including the time scale of 
properties of low mass stars (lithium abundance, rotation, stellar 
activity), pre-main sequence stars, and internal structure and evolutionary 
models.

\end{document}